\documentclass[twocolumn,preprintnumbers,prl,superscriptaddress,amsmath,amssymb,amsfonts]{revtex4}

\usepackage{graphicx}  % needed for figures
\usepackage{color}
\usepackage[colorlinks,bookmarks=true,citecolor=blue,linkcolor=blue,urlcolor=blue, breaklinks=true]{hyperref}
\usepackage{bm}
\setcitestyle{super}
\begin{document}

%\preprint{}
\title{Evidence for the first-order topological phase transition in a Kitaev spin liquid candidate $\alpha$-RuCl$ _3$}

\author{S. Suetsugu}
\affiliation{Department of Physics, Kyoto University, Kyoto 606-8502, Japan}
\author{Y. Ukai}
\affiliation{Department of Physics, Kyoto University, Kyoto 606-8502, Japan}
\author{M. Shimomura}
\affiliation{Department of Physics, Kyoto University, Kyoto 606-8502, Japan}
\author{M. Kamimura}
\affiliation{Department of Physics, Kyoto University, Kyoto 606-8502, Japan}
\author{T. Asaba}
\affiliation{Department of Physics, Kyoto University, Kyoto 606-8502, Japan}
\author{Y. Kasahara}
\affiliation{Department of Physics, Kyoto University, Kyoto 606-8502, Japan}
\author{N. Kurita}
\affiliation{Department of Physics, Tokyo Institute of Technology, Tokyo 152-8551, Japan}
\author{H. Tanaka}
\affiliation{Department of Physics, Tokyo Institute of Technology, Tokyo 152-8551, Japan}
\author{T. Shibauchi}
\affiliation{Department of Advanced Materials Science, University of Tokyo, Chiba 277-8561, Japan}
\author{J. Nasu}
\affiliation{Department of Physics, Tohoku University, Sendai 980-8578, Japan}
\author{Y. Motome}
\affiliation{Department of Applied Physics, University of Tokyo, Tokyo 113-8656, Japan.}
\author{Y. Matsuda}
\affiliation{Department of Physics, Kyoto University, Kyoto 606-8502, Japan}

\date{\today}

% \begin{abstract}
% \end{abstract}

%\keywords{}

\maketitle

\textbf{The Kitaev quantum spin liquid (QSL) \cite{kitaev2006anyons} on the two-dimensional honeycomb lattice epitomizes an entangled topological state, where the spins fractionalize into Majorana fermions. This state has aroused tremendous interest because it harbors non-Abelian anyon excitations. The half-integer quantized thermal Hall (HIQTH) conductance observed in $\alpha$-RuCl$_3$ \cite{kasahara2018majorana,yokoi2021half,PhysRevB.102.220404,bruin2021robustness} is a key signature of these excitations.  However, the fate of this topologically nontrivial state at intense fields remains largely elusive. Here, we report the thermal conductivity $\kappa$ and specific heat $C$ of $\alpha$-RuCl$_3$ with in-plane magnetic fields $H$. For the field direction perpendicular to the Ru-Ru bond, where the HIQTH effect is observed, we find a discontinuous jump in $\kappa(H)$ and a peak anomaly in $C(H)$ at $\mu_0H^*\approx11$\,T, evidencing a weak first-order phase transition. Remarkably, the HIQTH effect vanishes close to $H^*$ \cite{yokoi2021half}. Furthermore, we find that the spin-fractionalization feature is retained well above $H^\ast$. These imply the emergence of the phase transition that separates two QSL phases with distinct topological properties.
}

\begin{figure}[b]
	\includegraphics[clip,width=8cm]{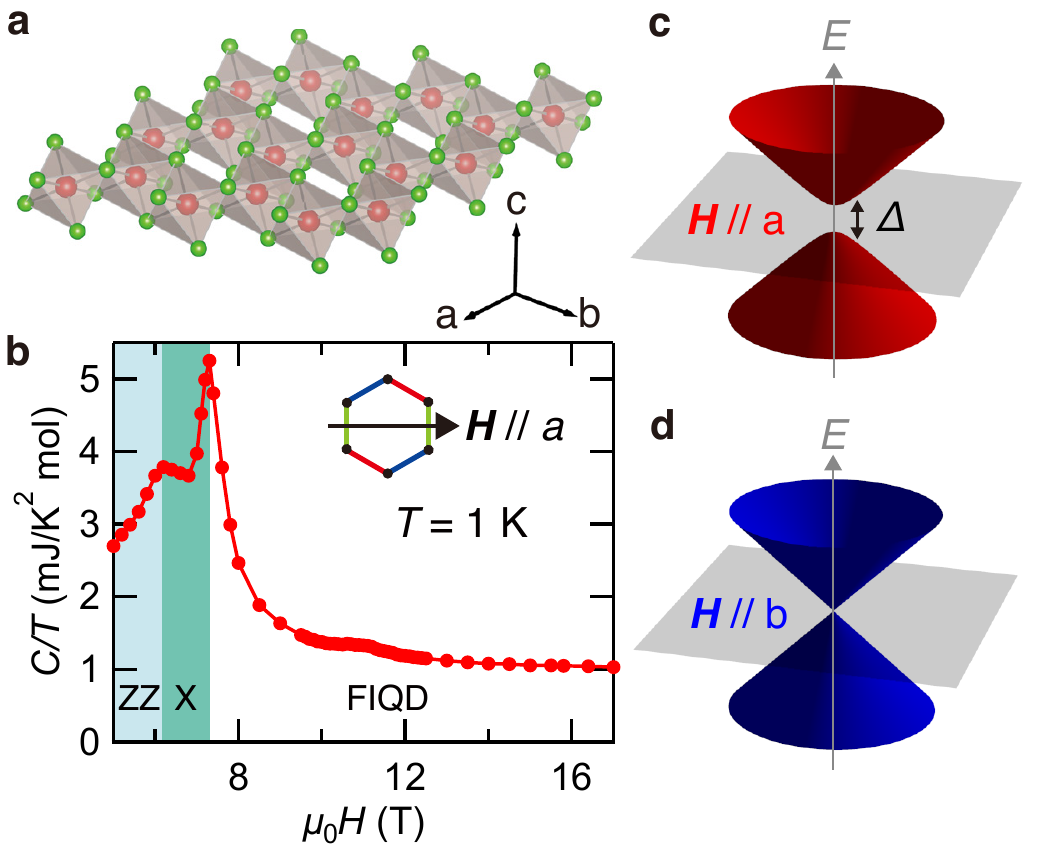}
	\caption{\textbf{Crystal structure, phase diagram and dispersion of Majorana bands in $\alpha$-RuCl$ _3$.}
		{\bf a}, Crystal structure of $\alpha$-RuCl$ _3$. Red and green spheres represent Ru$ ^{3+}$ and Cl$ ^{-}$ ions, respectively. The crystal $a$-axis ($b$-axis) is perpendicular (parallel) to the Ru-Ru bond direction in the honeycomb plane. {\bf b}, Magnetic field dependence of the specific heat at 1\,K ({\boldmath $H$}$\parallel a$). Peaks at 6.2\,T and 7.3\,T correspond to the phase transitions from the zigzag (ZZ) AFM phase to the so-called ``X" AFM phase and then to the field-induced quantum disordered (FIQD) state, respectively. {\bf c, d}, Band dispersion of itinerant Majorana fermions in magnetic fields. The gap $\Delta$ opens for ${\bm H}\parallel a$ ({\bf c}), while the gapless Dirac cone appears for ${\bm H}\parallel b$ ({\bf d}).
		\label{fig:RuCl3}
	}
\end{figure}

In QSLs, frustration and quantum fluctuations prevent the interacting spins from developing long-range magnetic order even at zero temperature \cite{balents2010spin}. QSLs are fascinating states of matter, accompanied by emergent topological order and fractionalized excitations. Among QSLs, the spin-1/2 Kitaev model on a two-dimensional (2D) honeycomb lattice \cite{kitaev2006anyons} offers an exactly solvable example. This model involves a highly entangled state resulting from bond-dependent Ising interactions that act as an exchange frustration. This leads to a topological QSL ground state that displays the fractionalization of spins into itinerant Majorana fermions and massive (gapped) $Z_2$ fluxes (visons). In magnetic fields, these emergent excitations form non-Abelian anyons that can be employed to perform fault resilient quantum information processing. The spin-orbit Mott insulator $\alpha$-RuCl$_3$ \cite{PhysRevB.90.041112} with a honeycomb lattice of Ru$^{3+}$ ions stacked along the $c$-axis direction (Fig.\,\ref{fig:RuCl3}a) is drawing much attention as a promising Kitaev QSL candidate \cite{PhysRevLett.102.017205,takagi2019kitaev}. The predominance of the Kitaev interaction in this compound has been suggested by various experiments and calculations \cite{motome2020hunting,winter2017models}. In fact, unusual magnetic continuum excitations revealed by Raman \cite{PhysRevLett.114.147201,nasu2016fermionic} and inelastic neutron scattering (INS) \cite{banerjee2016proximate,do2017majorana,banerjee2018excitations,PhysRevB.100.060405} and the fractional entropy reported by the specific heat measurements \cite{do2017majorana,PhysRevB.99.094415} have been suggested to be fingerprints of the spin fractionalization into Majorana fermions. At zero field, $\alpha$-RuCl$ _3$ undergoes antiferromagnetic (AFM) transition at around 7\,K \cite{PhysRevB.92.235119} due to the presence of non-Kitaev type interactions, arising from Heisenberg and off-diagonal interactions. However, in-plane magnetic fields suppress the AFM order, leading to a field-induced quantum disordered (FIQD) state above $\sim 7$\,T \cite{banerjee2018excitations,PhysRevB.103.174417} (Fig.\,1b).

A fundamental question is as to whether a distinct intermediate phase is present between the AFM ordered phase and a fully spin-polarized phase in the strong field limit. Despite intensive research efforts, however, the exact nature of the FIQD state has been highly controversial. It has been reported that in the  FIQD state, the thermal Hall conductivity per 2D layer exhibits a quantized plateau at half-integer values \cite{kasahara2018majorana,yokoi2021half,PhysRevB.102.220404,bruin2021robustness}, $\kappa_{xy}^{2D}=\frac{1}{2}C_h\cdot K_0$, where $K_0=(\pi^2/3)(k_B^2/h)T$ is the quantum thermal conductance and $C_h=\pm 1$ is the topological Chern number. The observation of the HIQTH effect demonstrates the emergence of non-Abelian anyons and chiral edge modes of the Majorana fermions. In addition, the observed field-angular variation of the HIQTH conductance \cite{yokoi2021half} has the same sign structure as the Chern number of itinerant Majorana bands. In particular, the HIQTH conductance is observed even when the field is applied within the 2D plane; $C_h=-1$ for antibond field direction ({\boldmath $H$} $\parallel a$), while $\kappa_{xy}^{2D}=0$, i.e. $C_h=0$, for bond direction ({\boldmath $H$}  $\parallel b$).

Furthermore, it has been reported that in the FIQD state, the magnetic contribution to specific heat exhibits an exponential behavior \cite{tanaka2020thermodynamic}, $C_M\propto \exp(\Delta/k_BT)$ for  {\boldmath $H$} $\parallel a$, indicating the formation of an energy gap $\Delta$ (Fig.\,\ref{fig:RuCl3}c), and $\Delta$ increases approximately as $\Delta\propto H^3$. In contrast, $C_M/T$ increases in proportion to $T$ for {\boldmath $H$} $\parallel b$, indicating the itinerant Majorana quasiparticle band with a gapless Dirac-cone-like dispersion (Fig.\,\ref{fig:RuCl3}d). These results provide strong support that the non-Abelian topological order persists even in the presence of non-Kitaev interactions in $\alpha$-RuCl$_3$. Very recently, it has been reported that $\kappa(H)$ exhibits periodic oscillations as a function of $1/H$ \cite{czajka2021oscillations}, similar to quantum oscillations of metals. Although the HIQTH effect is not observed in ref.\,\citenum{czajka2021oscillations}, the presence of an emergent exotic phase has been suggested. A finite field regime of intermediate QSL phase within the FIQD state for {\boldmath $H$} $\parallel a$ has also been suggested by the INS and magnetocaloric measurements \cite{PhysRevB.100.060405}.

On the other hand, no additional transitions have been suggested in the FIQD state by several experiments \cite{PhysRevB.103.054440,modic2021scale,PhysRevLett.125.037202}. Moreover, while the magnetic excitations in magnetic fields have been interpreted in terms of the Majorana-magnon crossover phenomena \cite{PhysRevB.101.100408}, they have also been discussed by the strong magnon damping \cite{winter2017breakdown,PhysRevResearch.2.033011}.

To uncover the presence of the intermediate topological phase, a key challenge is to elucidate a phase transition inside the FIQD state. Moreover, given the presence of the transition, it is also crucially important to clarify how the topologically non-trivial phase evolves at intense magnetic fields, in particular, whether the high-field state is a simple spin-polarized state where the spin-fractionalization does not play a role.  Here, in order to unveil the nature of the FIQD state, we measured the longitudinal thermal conductivity $\kappa$ ($\kappa_{xx}$) and specific heat $C$ down to 200\,mK on the high-quality $\alpha$-RuCl$_3$ single crystal (see also Fig.\,S1), in which the HIQTH effect is observed \cite{yokoi2021half}. While $\kappa$ is sensitive exclusively to low-energy itinerant excitations, $C$ contains contributions from both localized and itinerant excitations.

\begin{figure}
\includegraphics[clip,width=8cm]{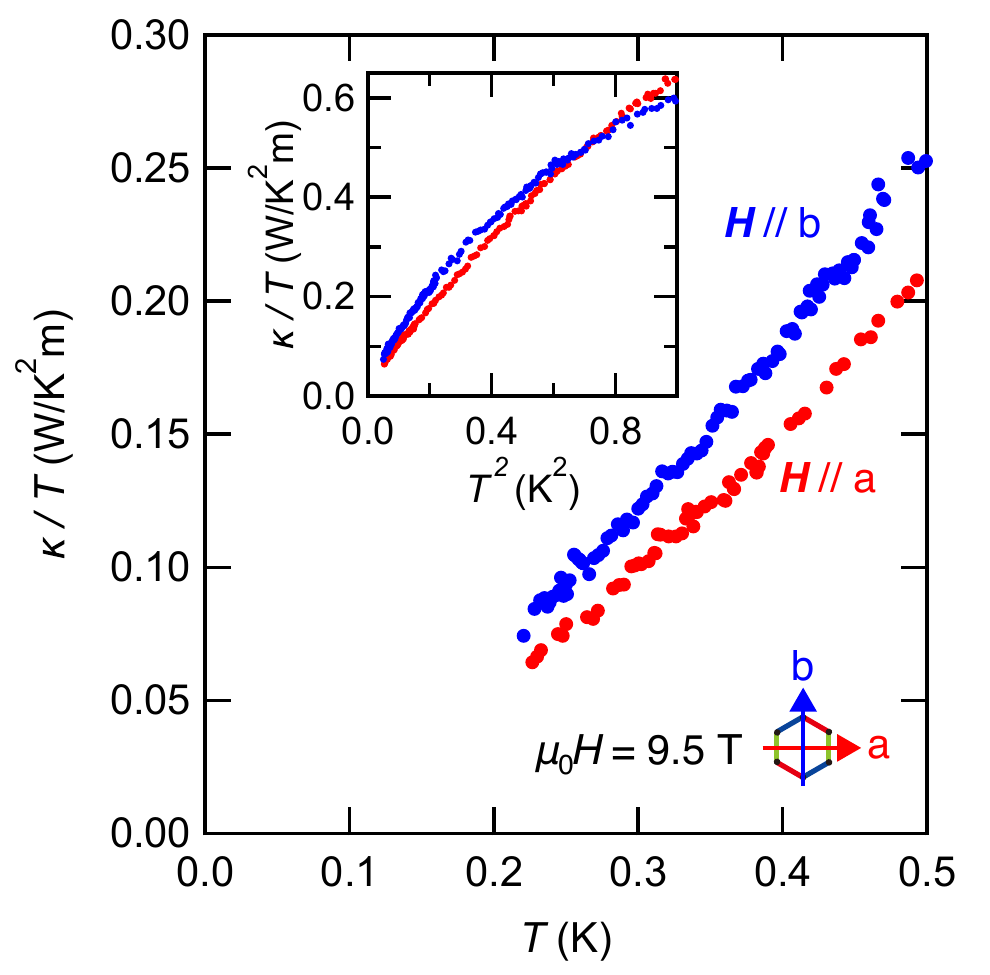}
\caption{\textbf{Temperature dependence of thermal conductivity in $\alpha$-RuCl$ _3$.}
	Thermal conductivity divided by temperature, $\kappa/T$, at very low temperatures plotted as a function of $T$ (main panel) for $\bm{H}\parallel a$ (red) and $\bm{H}\parallel b$ (blue) at 9.5\,T in the FIQD state. The inset plots $\kappa/T$ as a function of $T^2$.
\label{fig:C}
}
\end{figure}

Figure\,1b shows the $H$-dependence of $C$ for ${\bm H} \parallel a$. The FIQD state appears above 7.3\,T. Figure\,2 depicts $\kappa/T$\,vs $T$ (thermal current $\bm{j} \parallel a$) at low temperatures for {\boldmath $H$}$\parallel a$ and {\boldmath $H$}$\parallel b$ at 9.5\,T in the FIQD state. If we extrapolate $\kappa/T$ to $T=0$ simply assuming $T$-linear dependence, $\kappa/T$ has a negative intercept for both field directions. The inset of Fig.\,2 plots $\kappa/T$\,vs $T^2$. The residual linear term of the thermal conductivity, $\kappa/T(T\rightarrow 0)$, is negligibly small for both directions. These results indicate the absence of finite density of states of gapless itinerant excitations, which is consistent with the absence of $\gamma$-term in the specific heat \cite{tanaka2020thermodynamic}.

\begin{figure}
	\includegraphics[clip,width=7cm]{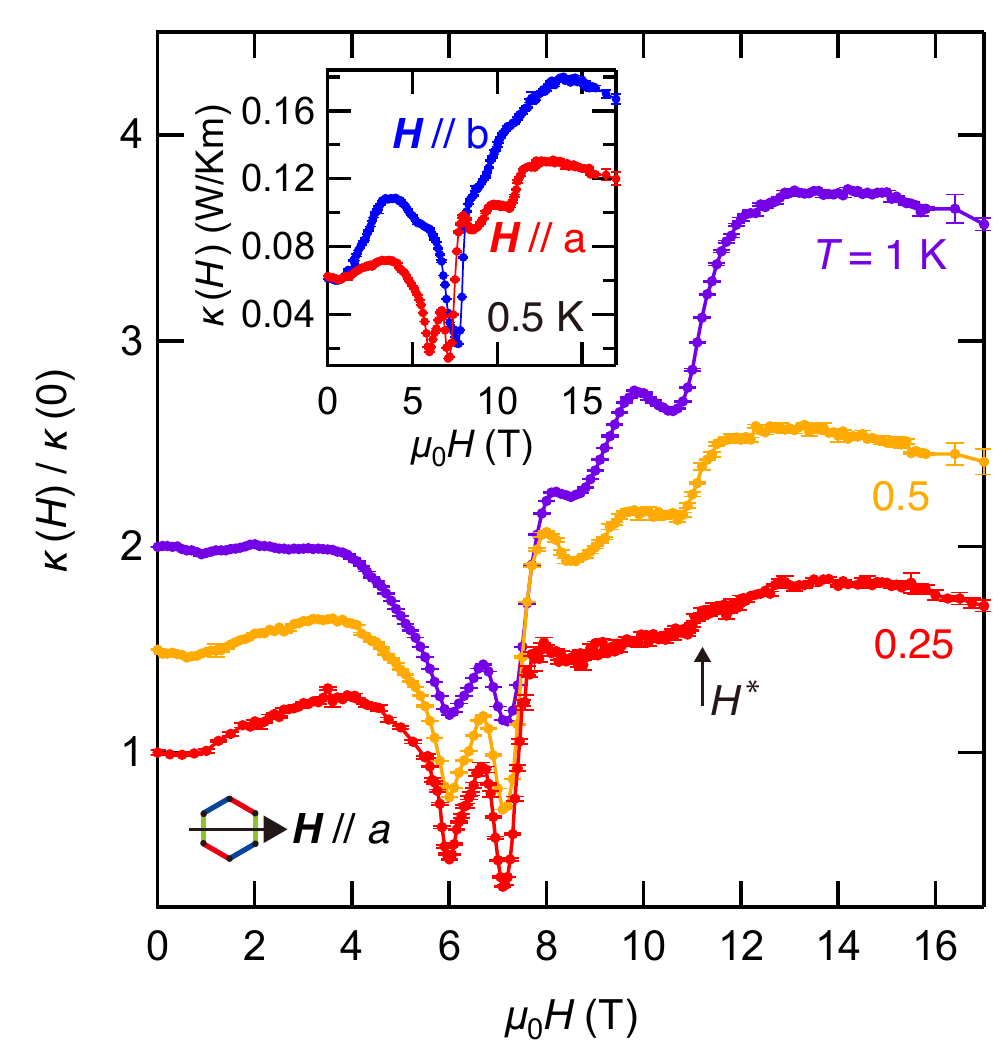}
	\caption{\textbf{Magnetic field dependence of thermal conductivity.}
		Field dependence of $\kappa$ normalized by its zero-field value for $\bm{H} \parallel a$ at 0.25, 0.5, and 1\,K. The data at 0.5 and 1\,K are vertically shifted for clarity. The sharp minima at 6.0\,T and 7.2\,T correspond to the phase transitions from the zigzag (ZZ) AFM phase to the  X-phase and then to the FIQD state, respectively. The first-order phase transition occurs at $H^*$ indicated by the arrow, as revealed by a discontinuous jump of $\kappa(H)$ at low temperatures. The inset shows the $H$-dependence of $\kappa$ at 0.5 K for $\bm{H}\parallel a$ and $\bm{H}\parallel b$. For ${\bm H}\parallel b$, $\kappa(H)$ exhibits a single minimum at $\mu_0H=7.7$\,T. This minimum is slightly broader than those for ${\bm H}\parallel a$, which is due to the presence of the X-phase in a very narrow field range.}
\end{figure}

It has been reported that for {\boldmath $H$}$\parallel a$ at 9.5\,T, the Majorana band gap is $\sim$10\,K \cite{tanaka2020thermodynamic}, while the magnon gap exceeds 20\,K \cite{PhysRevB.100.060405}. Therefore, Majorana quasiparticles hardly carry the heat and hence $\kappa$ is dominated by the phonon contribution, $\kappa\approx \kappa_{ph}$. However, in the presence of coupling between emergent excitations and phonons, $\kappa_{ph}$ is influenced through the scattering with the excitations even at low temperatures. It should be stressed that the coupling of Majorana fermions and phonons is proposed to be required for the observation of the HIQTH effect \cite{PhysRevX.8.031032,PhysRevLett.121.147201}. As shown in Fig.\,2, $\kappa/T$ for $\bm{H}\parallel b$ is larger than $\bm{H}\parallel a$ at very low temperatures. This indicates the presence of itinerant quasiparticles that conduct heat; $\kappa=\kappa_{ph}+\kappa_{qp}$ for $\bm{H} \parallel b$. Although lower temperature measurements are required to fully unveil the low-energy excitations, the finite $\kappa_{qp}$-term, along with $\kappa_{qp}/T(T\rightarrow 0)=0$ obtained from $\kappa/T(T\rightarrow 0)=0$, supports the presence of gapless Dirac cones, as reported by specific heat \cite{tanaka2020thermodynamic}.
At high temperatures, $\kappa$ for ${\bm H}\parallel b$ becomes lower than ${\bm H}\parallel a$. This can be explained by considering that $\kappa_{ph}$ for ${\bm H}\parallel b$ is suppressed at high temperatures due to the rapidly increasing number of Majorana fermions excited in the Dirac cones, which decrease the phonon mean free path.

\begin{figure*}
	\includegraphics[clip,width=16cm]{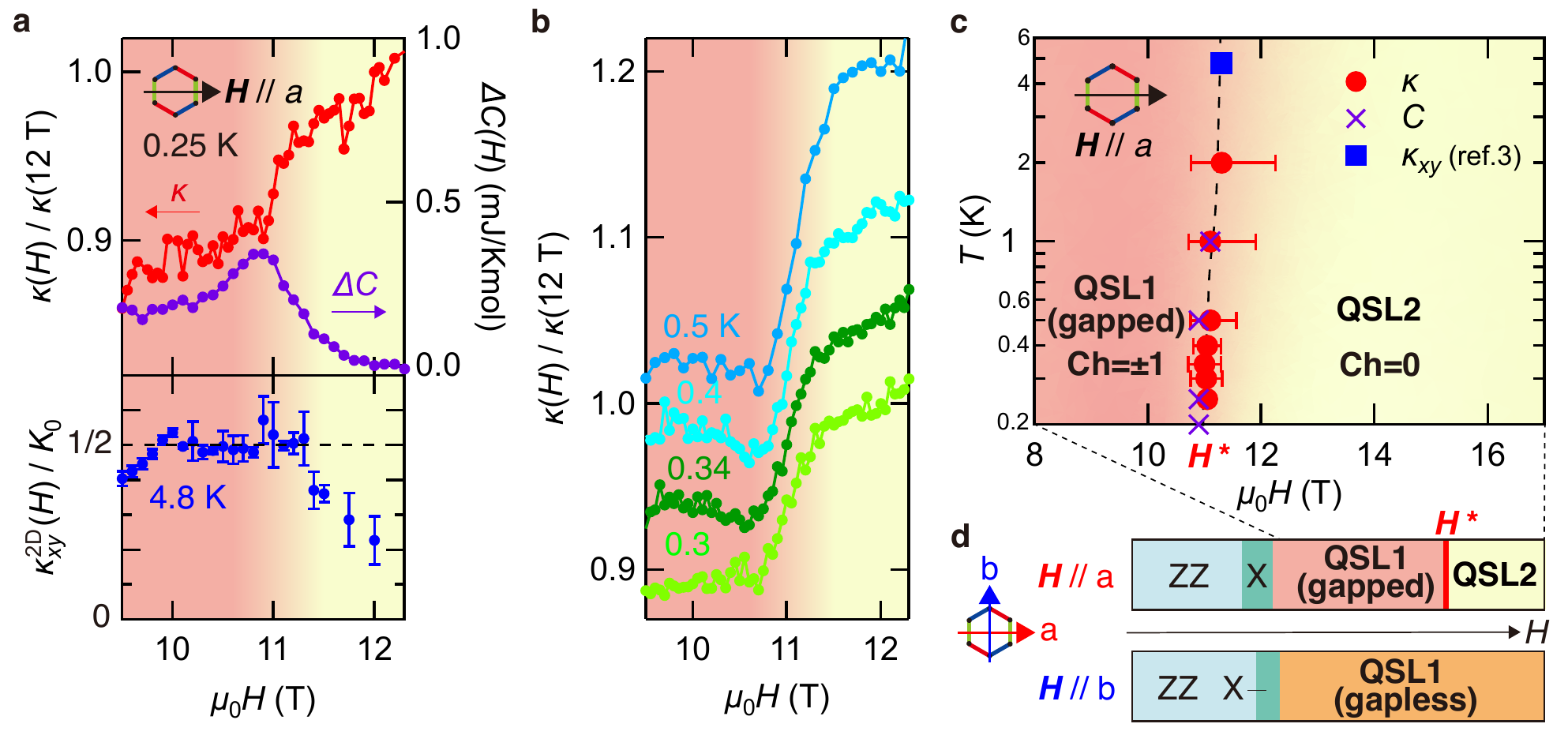}
	\caption{\textbf{First-order topological phase transition for $\bm{H}\parallel a$ and phase diagram.}
		{\bf a}, The upper panel shows the field dependence of the thermal conductivity normalized by the value at $\mu_0H=12$\,T and the relative change of the specific heat $\Delta C(H)=C(H)-C(12\,{\rm T})$ at 0.25\,K for $\bm{H}\parallel a$. At $\mu_0H^\ast\approx 11$\,T, $\kappa(H)$ exhibits a discontinuous jump and $\Delta C(H)$ shows a peak. The lower panel shows the thermal Hall conductivity per 2D layer $\kappa_{xy}^{2D}$ normalized by the quantum thermal conductance $K_0$ (taken from ref.\,\citenum{yokoi2021half}). {\bf b}, The field dependence of the normalized thermal conductivity, $\kappa(H)/\kappa(12\,{\rm T})$, at higher temperatures. The data for $T \geq$ 0.34\,K are vertically shifted for clarity. {\bf c}, $H$-$T$ phase diagram of $\alpha$-RuCl$_3$ for ${\bm H}\parallel a$ in the FIQD state determined by the jump of the thermal conductivity (red circles), deviation from the half-integer plateau of the thermal Hall conductivity (blue square), and the peak of the specific heat (violet crosses) (see also Supplementary Figs.\,S3 and S4). For $T\geq$ 0.5\,K, where the kinks in the thermal conductivity are smeared out, the red points represent the inflection points in the jump. The error bars are determined by the starting and end points of the jump. The black dashed line is a guide to the eyes. The gapped Kitaev quantum spin liquid (QSL1) characterized by topological Chern number $(\pm 1)$ undergoes a weak first-order transition.
		Above the transition, a quantum spin liquid state characterized by zero Chern number (QSL2) appears. The spin-fractionalization may be still important in the QSL2 phase. {\bf d}, Schematic phase diagrams for $\bm{H}\parallel a$ (top) and $\bm{H}\parallel b$ (bottom) at $T=0$\,K. The zigzag (ZZ) state is suppressed by the field and the magnetically-ordered phase, indicated by ``X", exists in a narrow field range. In the FIQD state, the first-order phase transition separates two QSL phases (QSL1 and QSL2) with distinct topological properties for $\bm{H}\parallel a$, while gapless QSL state (QSL1) with the Dirac-cone dispersion of the Majorana band appears and persists with no phase transition for $\bm{H}\parallel b$.}
\end{figure*}

As shown in Fig.\,3 and its inset,  $\kappa(H)$ exhibits sharp minima at the AFM transitions (6.0 and 7.2\,T for ${\bm H}\parallel a$ and 7.7\,T for ${\bm H}\parallel b$). These sharp minima are caused by the rapid suppression of $\kappa_{ph}$ owing to the enhancement of phonon scattering by magnons, whose gap vanishes as $H$ approaches the AFM phase boundaries. In contrast to ref.\,\citenum{czajka2021oscillations}, the oscillation behavior has not been observed in the AFM ordered state. Moreover, in the FIQD state, oscillation-like $H$-dependence of $\kappa(H)$ disappears at low temperatures. We note that these are consistent with the absence of the oscillating behavior in the specific heat $C(H)$.

For ${\bm H}\parallel a$, $\kappa(H)$ exhibits a peak around 8\,T on entering the FIQD state from the ordered phase. Such an overshoot behavior is often observed in low-dimensional magnetic systems \cite{PhysRevB.87.224413,li2020possible} and may be related to the short-range AFM order. For ${\bm H}\parallel b$, $\kappa(H)$ exhibits a shoulder structure just above the AFM order. For ${\bm H}\parallel a$, $\kappa(H)$ shows a broad peak around $9.5$\,T at $T$ = 1\,K. This peak is less pronounced at 0.5\,K and almost vanishes at 0.25\,K. Such a peak structure is not observed for ${\bm H}\parallel b$. As we will discuss later, the observed peak structure can be attributed to the suppression of $\kappa(H)$ above $\sim$ 10\,T.

One of the most notable features in the FIQD state is the anomaly found at $\mu_0H^*\approx11$\,T (arrow in Fig.\,3), where $\kappa(H)$ increases steeply at 0.5 and 1\,K. As shown in Fig.\,4a, $\kappa(H)$ at 0.25\,K exhibits a discontinuous jump at $H^*$, which is characterized by distinct kinks at the starting and end points and a very narrow transition width of $\Delta H/H^*\leq 10^{-2}$. The jump of $\kappa(H)$ indicates that a first-order phase transition takes place at $H^*$. The thermodynamic evidence of the phase transition is further provided by the specific heat, which exhibits a distinct peak at $H^*$ (Fig.\,4a). The observed peak in $C(H)$ is relatively broad between $\sim$10 and $\sim12$\,T, indicating the presence of the fluctuation regime. These results suggest that the transition is weak first-order. As the discontinuous jump is absent for ${\bm H}\parallel b$, it is natural to consider that the first-order transition is related to the topological nature of Kitaev QSL. To confirm this, we also plot the $H$-dependence of $\kappa_{xy}^{2D}$ measured in the same crystal at 4.5\,K \cite{yokoi2021half} in Fig.\,4a. Remarkably, the HIQTH  plateau disappears around $H^*$. These results lead us to conclude the first-order topological phase transition in the FIQD state. A tiny specific heat anomaly has been reported for $\bm{H}\parallel a$ at 0.7\,K around 10-10.5\,T in ref.\,\citenum{tanaka2020thermodynamic}, which is likely to have the same origin as that at $H^\ast$. The difference between the anomaly fields may be due to crystal-dependent interaction parameters. We note that such an anomaly is absent for $\bm{H}\parallel b$ \cite{tanaka2020thermodynamic}, which is consistent with the thermal conductivity results.

As $\kappa\approx\kappa_{ph}$ for ${\bm H}\parallel a$, the jump of $\kappa(H)$ at $H^*$ is due to the jump of the phonon mean free path $\ell_{ph}$, which is caused by a sharp change of the phonon-quasiparticle scattering rate. This scenario is supported by the fact that the magnitude of the jump of $\kappa(H)$ at $H^*$ is largely suppressed at low temperatures and eventually the jump is not discernible at 200\,mK (see Fig.\,S2), where $\ell_{ph}$ becomes comparable to the sample size. As shown in Fig.\,4b, the kinks of $\kappa(H)$ at the transition are preserved at 0.3\,K, but become obscure at higher temperatures. 
Above 0.4\,K, $\kappa(H)$ peaks around $10$\,T, followed by a minimum just below $H^*$. The suppression of $\kappa(H)$ above $\sim10$\,T could be attributed to the fluctuations, which are observed in $C(H)$. As $C$ contains contributions of both localized and itinerant excitations, the peak of $C(H)$ at low temperatures arises from localized quasiparticle contributions, most likely localized Majorana fermions.

As shown in the inset of Fig.\,3,  $\kappa$ is still largely anisotropic even at 17\,T; $\kappa({\bm H}\parallel a)<\kappa({\bm H}\parallel b)$. Moreover, $\kappa(H)$ decreases with $H$ above $\sim 14$\,T for both field directions. These results imply that even well above $H^*$, the system is not in a simple partially spin-polarized state, where magnetic (spin-wave) excitations are gapped with a Zeeman gap $\mu_BH$, as explained below. In such a state, because of the large Zeeman gap, spin-phonon scattering is absent and, in addition, spins do not carry the heat. Then, $\kappa$ is expected to be dominated by isotropic and $H$-independent phonon contribution without spin scattering.
Therefore, the observed anisotropic and $H$-dependent $\kappa(H)$ indicates that the presence of low-energy quasiparticle excitations that scatter phonons. Although the Majorana picture in the Kitaev model is expected to break down in intense fields, the present results suggest that the spin-fractionalization still plays an important role and the QSL state is retained even at 17\,T. Note that recent theoretical studies suggest the field-induced paramagnetic phase distinct from the simple spin-polarized state exists in the vicinity of the Kitaev QSL phase \cite{lee2020magnetic,PhysRevResearch.3.023189}.

In Fig.\,4c, we illustrate the phase diagram in the FIQD state determined by the specific heat and thermal transport measurements for ${\bm H}\parallel a$. The weak first-order phase transition separates two QSL phases (QSL1 and QSL2) with distinct topological properties. Using the Clausius-Clapeyron equation, the almost $T$-independent transition line leads to an entropy jump $\Delta S \propto dH^\ast / dT \approx 0$. Therefore, the latent heat at the first-order transition is expected to be very small. As shown in the schematic phase diagrams in Fig.\,4d, the high-field QSL2 phase appears above the first-order transition for ${\bm H}\parallel a$, while the gapless QSL state with the Dirac-cones of the Majorana bands persists without such a phase boundary for ${\bm H}\parallel b$.

Our results imply the emergence of the high-field QSL2 phase with topological properties different from the QSL1 phase. Although the appearance of a nematic phase has been reported above $\sim H^*$ \cite{tanaka2020thermodynamic} and discussed in the context of a possible toric code phase \cite{PhysRevResearch.3.023189}, its nature remains highly elusive. Exploring the high-field phase and the first-order topological transition can provide a key to understanding the mechanism of emergence and destruction of the non-Abelian topological order, which is a pivotal issue of error tolerant quantum computation.

\noindent
{\bf Methods}\\
{\bf Single crystal growth and characterization.}
High-quality single crystals of $\alpha$-RuCl$ _3$ were grown by a vertical Bridgman method as described in ref.\,\citenum{PhysRevB.91.094422}. For specific heat and thermal conductivity measurements, crystal \#3 in ref.\,\citenum{yokoi2021half}, in which the HIQTH effect is observed, was used. The direction of the crystal axes was determined by Laue X-ray back reflection measurements. The size of the crystal is 2150 $\mathrm{\mu m}$ (length) $\times$ (1000 $\pm$ 10) $\mathrm{\mu m}$ (width) $\times$ (20 $\pm$ 0.3) $\mathrm{\mu m}$ (thickness). Fig.\,S1 shows the temperature dependence of specific heat $C$ of this single crystal at 0\,T. A clear peak anomaly is observed at 7.3\,K.
\\
{\bf Specific heat and thermal conductivity.}
Specific heat $C$ was measured by a long-relaxation time method by applying magnetic field $H$ along the crystal $a$-axis ($\bm{H} \parallel a$) as described in refs.\,\citenum{PhysRevB.63.094508,PhysRevLett.99.057001}. Thermal conductivity $\kappa$ was measured by the standard steady-state method by applying thermal current $\bm{j} \parallel a$. Three gold wires were attached by silver paste to serve heat links to a 43-$\mathrm{k\Omega}$ chip as a heater and two thermometers. One end of the crystal was glued to a silver plate as a heat bath using silver paste. $\alpha$-RuCl$ _3$ has three equivalent $a$-axes because of the threefold rotational symmetry of the crystal structure. In the thermal conductivity measurements for $\bm{H} \parallel a$, the field was applied along the same $a$-axis as in the specific heat measurements.

\bibliography{ref.bib}

\noindent
{\bf Acknowledgments}\\
We thank S. Fujimoto, M. G. Yamada, Y. Mizukami, and K. Hashimoto for insightful discussions. This work is supported by Grants-in-Aid for Scientific Research (KAKENHI) (Nos. 18H05227, 18H03680, 18H01180, 21K13881) and on Innovative Areas ``Quantum Liquid Crystals'' (Nos. JP19H05824 and JP19H05825) from the Japan Society for the Promotion of Science, and JST CREST (JPMJCR19T5).

\noindent
{\bf Author contributions}\\
N.K. and H.T. synthesized the high-quality single crystals. S.S., Y.U., M.S., and M.K. performed specific heat experiments. S.S., Y.U., and M.S. performed thermal conductivity experiments. S.S., Y.U., M.S., M.K., Y.K., T.A., and Y.Ma. analyzed the data. All authors discussed the results. S.S., T.S., J.N., Y.Mo., and Y.Ma. prepared the manuscript with inputs from all authors.

\noindent
{\bf Additional information}\\
The data that support the findings of this study are available on request from the corresponding author.

\noindent
{\bf Competing financial interests}\\
The authors declare no competing financial interests.

\end{document}